\documentclass[10pt]{ismd08}
\usepackage{graphicx}
\usepackage{amssymb}
\begin{document}
\title{On factorization scheme suitable for NLO Monte Carlo event generators}
\author{Karel Kolar}
\institute{Institute of Physics, Academy of Sciences of the Czech Republic, Prague}
\maketitle
\begin{abstract}
The choice of a factorization scheme suitable for Monte Carlo simulations of NLO initial
state parton showers is discussed in this contribution.
\end{abstract}

\section{Introduction}
Generating initial state parton showers (IPS) in hadronic collisions at the NLO accuracy is a task
for which no satisfactory solution has so far been found. An attempt at solving this problem which
is presented in this contribution is based on an exploitation of the freedom in the choice of
the factorization scheme.

As a simple illustration, consider a non--singlet nucleon structure function $F_{\rm NS} \!\left(
x, Q^2 \right)$. Its Mellin moments are given as the product
\begin{equation}
  F_{\rm NS} \!\left( n, Q^2 \right) = C_{\rm NS} \!\left( n, {Q\over M}, {\rm FS} \right)
  q_{\rm NS} \!\left( n, M, {\rm FS} \right), \label{eq1}
\end{equation}
where $C_{\rm NS} (n,Q/M,{\rm FS})$ stands for Mellin moments of the corresponding coefficient
function and $q_{\rm NS} (n,M,{\rm FS})$ represents Mellin moments of the relevant non--singlet
parton distribution function. Both $C_{\rm NS} (n,Q/M,{\rm FS})$ and $q_{\rm NS} (n,M,{\rm FS})$
depend on a particular factorization scheme $\rm FS$ and on a factorization scale $M$, however,
their product (\ref{eq1}) is independent of both of them. The coefficient function
$C_{\rm NS} (n,Q/M,{\rm FS})$ is calculable within the framework of perturbative QCD and can
thus be expanded in powers of the QCD couplant $a \equiv \alpha_{\rm s} / \pi$
\begin{equation}
  C_{\rm NS} \!\left( n, {Q\over M}, {\rm FS} \right) = \sum_{k=0}^{\infty} a^{k}(\mu, {\rm RS})
  \, C^{(k)}_{\rm NS} \!\left( n, {Q\over M}, {\rm FS}, \mu, {\rm RS} \right). \label{eq2}
\end{equation}
Although both the couplant $a(\mu, {\rm RS})$ and the coefficients $C^{(k)}_{\rm NS} (n,Q/M,{\rm FS},
\mu, {\rm RS})$ depend on a particular renormalization scheme $\rm RS$ and on a renormalization
scale $\mu$, which is in principle different from $M$, the series, if summed to all orders, is
independent of both the $\rm RS$ and $\mu$. The non--singlet parton distribution function
$q_{\rm NS} (n,M,{\rm FS})$ satisfies the evolution equation\footnote{From the relations
(\ref{eq3}) and (\ref{eq4}), we see that the non--singlet distribution function $q_{\rm NS}
(n,M,{\rm FS})$ also depends on the renormalization scheme in which the renormalized couplant
$a(M)$ is defined. This scheme can in principle be different from that used for the expansion
of the coefficient function $C_{\rm NS} (n,Q/M,{\rm FS})$, but usually these schemes are chosen
to be identical.}
\begin{equation}
  {{\rm d} q_{\rm NS} \!\left( n, M, {\rm FS} \right) \over {\rm d} \ln M} = a(M)\, P_{\rm NS} \!\left(
  n, M, {\rm FS} \right) q_{\rm NS} \!\left( n, M, {\rm FS} \right), \label{eq3}
\end{equation}
where the non--singlet splitting function $P_{\rm NS} (n,M,{\rm FS})$ can be expanded in powers of $a(M)$
\begin{equation}
  P_{\rm NS} \!\left( n, M, {\rm FS} \right) = \sum_{k=0}^{\infty} a^{k}(M)\, P^{(k)}_{\rm NS} \!\left(
  n, {\rm FS} \right). \label{eq4}
\end{equation}

In the next--to--leading order approximation, we retain only the first two terms in the expansions
(\ref{eq2}) and (\ref{eq4}):
\begin{eqnarray}
  C_{\rm NS} \!\left( n, {Q\over M}, {\rm FS} \right) & \!\! = \!\! & C^{(0)}_{\rm NS} (n) +
  a(\mu) \, C^{(1)}_{\rm NS} \!\left( n, {Q\over M}, {\rm FS} \right), \\
  P_{\rm NS} \!\left( n, M, {\rm FS} \right) & \!\! = \!\! & P^{(0)}_{\rm NS} (n) +
  a(M)\, P^{(1)}_{\rm NS} \!\left( n, {\rm FS} \right)
\end{eqnarray}
and the NLO couplant $a(\mu)$ obeys the differential equation
\begin{equation}
  {{\rm d} a(\mu) \over {\rm d}\ln\mu} = -b a^2(\mu) \Bigl( 1 + c a(\mu) \Bigr) .
\end{equation}
The LO terms $C^{(0)}_{\rm NS} (n)$ and $P^{(0)}_{\rm NS} (n)$ are universal --- independent of any
unphysical quantities such as renormalization and factorization scales and schemes. The NLO contributions
$C^{(1)}_{\rm NS} \!\left(n, Q/M, {\rm FS} \right)$ and $P^{(1)}_{\rm NS} \!\left( n, {\rm FS} \right)$
satisfy the following condition
\begin{equation}
  C^{(1)}_{\rm NS} \!\left( n, {Q\over M}, {\rm FS} \right) = C^{(0)}_{\rm NS}(n) \left[ \kappa (n) +
  P^{(0)}_{\rm NS}(n) \ln {Q\over M} + {1 \over b} P^{(1)}_{\rm NS} \!\left( n, {\rm FS} \right)
  \right], \label{eq8}
\end{equation}
where $\kappa (n)$ is a scale and scheme factorization invariant. The ambiguity related to the factorization
procedure is already at the NLO large, because the splitting function $P^{(1)}_{\rm NS} (x, {\rm FS})$
is completely arbitrary function --- for any function $f(x)$, there always exists such a factorization
scheme FS in which $P^{(1)}_{\rm NS} (x, {\rm FS}) = f(x)$. There are two prominent choices of the
splitting function $P^{(1)}_{\rm NS} (n, {\rm FS})$, which are in some sense opposite to each other.
In the first one, the splitting function $P^{(1)}_{\rm NS} (n, {\rm FS})$ is set equal to zero. For
this choice, which will be called the ZERO factorization scheme, the evolution of the distribution
function $q_{\rm NS} (n,M,{\rm FS})$ is formally identical to the LO one and all NLO corrections are
thus contained in the NLO coefficient function $C^{(1)}_{\rm NS} (n,Q/M,{\rm FS})$. The latter choice
consists in selecting the splitting function $P^{(1)}_{\rm NS} (n, {\rm FS})$ in such a way that the NLO
coefficient function $C^{(1)}_{\rm NS} (n,Q/M,{\rm FS})$ vanishes for $M = Q$ (see the equation (\ref{eq8})).
In this case, the relation between the structure function $F_{\rm NS} \!\left( x, Q^2 \right)$ and
the distribution function $q_{\rm NS} (n,M,{\rm FS})$ has the same form as in the LO and all NLO
corrections are included in the evolution of the distribution function $q_{\rm NS} (n,M,{\rm FS})$.
This type of choice is the essence of the so called DIS factorization scheme introduced in \cite{dis},
which is widely used in phenomenology. The factorization scheme dependence of NLO theoretical predictions
for physical quantities is studied, for instance, in \cite{soper}, where only factorization schemes
interpolating between the DIS and $\overline{\rm MS}$ factorization schemes are considered.

The factorization scheme specifies the way in which the so called collinear singularities, which
are contained in cross--sections at parton level, are absorbed into the dressed parton distribution
functions. Within the framework of dimensional regularization, the relation between the dressed
and bare distribution functions is given in the general case by the formula
\begin{eqnarray}
  D_i(x, M, {\rm FS}) &\!\!\!\! = \!\!\!\! & \sum_j \int_x^1 {{\rm d}y\over y} D_j^{(0)} \!\!\left(
  {x\over y} \right) \!\left[ \delta_{ij} \delta (1-y) + a(M)\! \left( {1\over\epsilon} A^{(11)}_{ij}(y)
  + A^{(10)}_{ij}(y) \!\right) + \right. \nonumber\\ & \!\!\!\! \!\!\!\! & + \left. a^2(M) \!\left( {1\over
  \epsilon^2} A^{(22)}_{ij}(y) + {1\over\epsilon} A^{(21)}_{ij}(y) + A^{(20)}_{ij}(y)\! \right) +
  \cdots \right] . \label{eq9}
\end{eqnarray}
The matrices $A^{(k0)}_{ij}(x)$ can be chosen arbitrarily and their choice fully specifies the factorization
scheme. The factorization scheme can also be specified by higher order splitting functions $P^{(k)}_{ij}(x,
{\rm FS})$, $k \ge 1$, which we can choose at will. The most widely used factorization scheme is the so called
$\overline{\rm MS}$ factorization scheme, which is defined by setting the matrices $A^{(k0)}_{ij}(x)$ equal
to zero\footnote{with the renormalized couplant $a(M)$ defined in the $\overline{\rm MS}$ renormalization
scheme} and is thus convenient for theoretical calculations.

At present time many QCD cross--sections at parton level are known at the NLO accuracy. However,
the algorithms that are used for their incorporation in Monte Carlo event generators attach to them
the IPS only at the LO accuracy because generating IPS at the NLO accuracy is very complicated in
the standard $\overline{\rm MS}$ factorization scheme. The reasons for that are basically two:
The NLO splitting functions no longer correspond to basic QCD vertices and the splitting functions
at the NLO approximation are negative for some $x$, which prevents us from using straightforward
probabilistic interpretation. Because the IPS induce the scale dependence of parton distribution
functions, it is inconsistent to attach the LO IPS to NLO QCD cross--sections, which include NLO
parton distribution functions. This deficiency could be removed by choosing the ZERO factorization
scheme, in which the NLO IPS are formally identical to the LO ones and all NLO corrections are
thus put into hard scattering cross--sections. The main advantage of this approach is the fact
that the existing algorithms for parton showering and for attaching parton showers to NLO
cross--sections need not be changed. The only step necessary to do is transforming parton level
cross--sections from the standard $\overline{\rm MS}$ factorization scheme to the ZERO factorization
scheme and determining parton distribution functions in the ZERO scheme.

\section{The transformation of hard scattering cross--sections}
In the case of a hadron collision, a cross--section $\sigma (P)$ (in general differential) depending on
observables $P$ is given by the formula
\begin{equation}
  \sigma (P) = \sum_{ij} \int_0^1 \!\int_0^1 {\rm d}x_1\,{\rm d}x_2\, D_{i/A} (x_1, M, {\rm FS}) \, D_{j/B}
  (x_2, M, {\rm FS}) \, \sigma_{ij} (x_1, x_2; P; M, {\rm FS}),
\end{equation}
where $D_{i/A} (x, M, {\rm FS})$ and $D_{i/B} (x, M, {\rm FS})$ are the parton distribution functions
of the colliding hadrons. The partonic cross--section $\sigma_{ij} (x_1, x_2; P; M, {\rm FS})$ can be
expanded in powers of the QCD couplant $a(\mu)$:
\begin{equation}
  \sigma_{ij} (x_1, x_2; P; M, {\rm FS}) = \sigma_{ij}^{(0)} (x_1, x_2; P) + a(\mu)\, \sigma_{ij}^{(1)}
  (x_1, x_2; P; M, {\rm FS}) + {\cal O} \!\left( a^2(\mu) \right) \! .
\end{equation}
The LO cross--section $\sigma_{ij}^{(0)} (x_1, x_2; P)$ is independent of the factorization scale and
scheme. The dependence of the NLO cross--section $\sigma_{ij}^{(1)} (x_1, x_2; P; M, {\rm FS})$ on the
factorization scale and scheme is determined by the formula
\begin{eqnarray}
  \lefteqn{\sigma_{ij}^{(1)} (x_1, x_2; P; M, {\rm FS}) =  \sigma_{ij}^{(1)} (x_1, x_2; P;
  M_0, \overline{\rm MS}) + \sum_{k} \int_0^1 {\rm d}y \biggl[ \sigma_{ik}^{(0)}
  (x_1, yx_2; P) \,\times \biggr. } \\ &\!\!\! \!\!\!& \times\!\left. \left( P^{(0)}_{kj}(y)
  \ln {M_0\over M} + T^{(1)}_{kj}(y, {\rm FS}) \!\right) + \sigma_{kj}^{(0)} (yx_1, x_2; P) \left(
  P^{(0)}_{ki}(y) \ln {M_0\over M} + T^{(1)}_{ki}(y, {\rm FS}) \!\right) \right] \! . \nonumber
\end{eqnarray}
What we need for the conversion to the factorization scheme FS is the knowledge of the corresponding
matrix function $T^{(1)}_{ij} (x, {\rm FS})$ \footnote{$T^{(1)}_{ij} (x, {\rm FS}) = -A^{(10)}_{ij}
(x, {\rm FS})$, where the matrix $A^{(10)}_{ij} (x, {\rm FS})$ is defined in the relation (\ref{eq9}).},
which is process independent. In the space of Mellin moments, the matrix $T^{(1)}_{ij} (n, {\rm FS})$
satisfies the following matrix equation:
\begin{equation}
  \left[ \mathbf{T}^{(1)}(n, {\rm FS}), \mathbf{P}^{(0)}(n) \right] - b \mathbf{T}^{(1)}(n, {\rm FS})
  = \mathbf{P}^{(1)}(n, \overline{\rm MS}) - \mathbf{P}^{(1)}(n, {\rm FS}) \, . \label{eq13}
\end{equation}

\section{The ZERO factorization scheme}
The solution of the preceding equation (\ref{eq13}) for the ZERO factorization scheme reads
\begin{equation}
\begin{array}{rlrl}
  T^{(1)}_{q_i q_j}(n) \!\!\!\! & = T^{(1)}_{\bar q_i \bar q_j}(n) = T^{(1)}_3(n) - {1\over b}
  P^{(1)}_{q_i q_j}(n, \overline{\rm MS}), \quad & T^{(1)}_{q_i G}(n) \!\!\!\! & = T^{(1)}_{\bar q_i G}
  (n) = T^{(1)}_1(n), \\[12pt]
  T^{(1)}_{q_i \bar q_j}(n) \!\!\!\! & = T^{(1)}_{q_i \bar q_j}(n) = T^{(1)}_3(n) - {1\over b}
  P^{(1)}_{q_i \bar q_j}(n, \overline{\rm MS}), & T^{(1)}_{G q_i}(n) \!\!\!\! & = T^{(1)}_{G \bar q_i}
  (n) = T^{(1)}_2(n), \\[12pt]
  T^{(1)}_{GG}(n) \!\!\!\! & = -{1\over b} P^{(1)}_{GG}(n, \overline{\rm MS}) - 2n_{\rm f} T^{(1)}_3(n), & &
\end{array}
\end{equation}
where the unknown functions $T^{(1)}_i(n)$ can be expressed in terms of the Mellin moments of the LO and
NLO splitting functions. The Mellin inversion of $T^{(1)}_i(n)$ has to be calculated numerically. This
was performed for three and four effectively massless flavours. The obtained results are very surprising
because for $x \lesssim 0.1$
\begin{equation}
  T^{(1)}_i(x) \approx C_i x^{-\xi} \quad {\rm with} \;\; \xi (n_{\rm f} = 3) \doteq 4.63 \;\; {\rm and}
  \;\; \xi (n_{\rm f} = 4) \doteq 3.85  \label{eq15}
\end{equation}
and the coefficients $C_i$ are so large that the functions $T^{(1)}_i(x)$ strongly dominate over
the NLO splitting functions $P^{(1)}_{kl}(x, \overline{\rm MS})$ in this region ($x \lesssim 0.1$). So
the question of applicability of the ZERO factorization scheme arises.

\begin{figure}[t]
\begin{center}
  \includegraphics[angle=90,width=\textwidth]{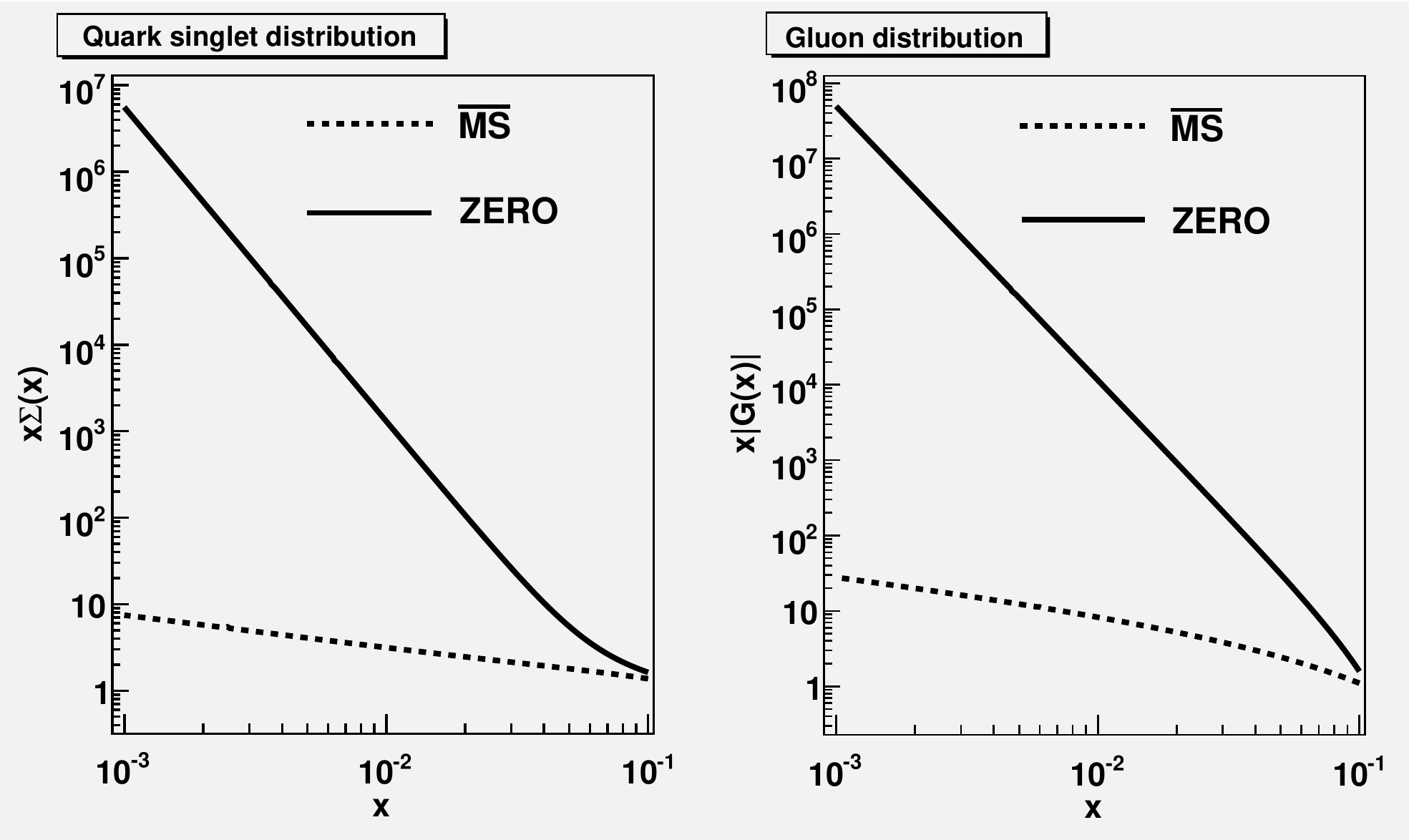}
  \caption{Comparison of the ZERO and $\overline{\rm MS}$ parton distributions at $M = 50\, {\rm GeV}$.
  The $\overline{\rm MS}$ distributions were obtained by evolving the starting distributions of
  the MRST98 set \cite{mrst} with the fixed number of active flavours $n_{\rm f} = 3$ (only light flavours
  are taken into account). The ZERO distributions were calculated from the $\overline{\rm MS}$ ones by using
  a numerical transformation method based on Mellin moments. Note that the gluon distributions are plotted
  in their absolute value because the ZERO gluon distribution is negative in the displayed region. The zero
  point where the ZERO gluon distribution changes the sign is close to $x = 0.1$.}
  \label{fig1}
\end{center}
\end{figure}

The parton distribution functions in the ZERO factorization scheme are plotted for $x \in (10^{-3}, 10^{-1})$
in Figure \ref{fig1}. In this region they behave like $x^{-\xi}$ with $\xi$ close to that in the relation
(\ref{eq15}). The ZERO factorization scheme can thus provide reasonable predictions only if large cancellation
between positive and negative values occurs in expressions for physical quantities. Hence, within the framework
of numerical computations, the ZERO factorization scheme can only be used in kinematic regions where
$x \gtrsim 0.1$. The only exception is its application in the non--singlet case, where the functions
$T^{(1)}_i(x)$ do not appear and no problems with large numbers occur.

\section{Summary and conclusion}
The ZERO factorization scheme is optimal for Monte Carlo simulations of NLO initial state parton showers.
However, because of the problems with large numbers, this scheme has too little range of applicability in
numerical calculations. The ZERO factorization scheme should thus be replaced by some ``almost ZERO''
factorization scheme which is sufficiently close to the ZERO factorization scheme and is free of problems
with large numbers. Searching for such a factorization scheme has already been started.

\begin{footnotesize}
\vspace{12pt}
\indent{\bf Acknowledgments.}\quad The author would like to thank J. Ch\'yla for careful reading of this
contribution and valuable suggestions. This work was supported by the projects LC527 of Ministry of
Education and AVOZ10100502 of the Academy of Sciences of the Czech Republic.

\end{footnotesize}

\end{document}